# Evaluating and Comparing Probability of Path Loss in DSDV, OLSR and DYMO at 802.11 and 802.11p

S. Wasiq[1], N. Javaid[1], M. Ilahi[1], R. D. Khan[2], U. Qasim[3], Z. A. Khan[4]

[1,2]COMSATS Institute of Information Technology, [1]Islamabad, [2]Wah Cant, Pakistan.
[3]University of Alberta, Alberta, Canada.
[4]Faculty of Engineering, Dalhousie University, Halifax, Canada.

**Abstract**

In this paper, we present path loss model for VANETs and simulate three routing protocols; Destination Sequenced Distance Vector (DSDV), Optimized Link State Routing (OLSR) and Dynamic MANET On-demand (DYMO) to evaluate and compare their performance using NS-2. The main contribution of this work is enhancement of existing techniques to achieve high efficiency of the underlying networks. After extensive simulations in NS-2, we conclude that DSDV best performs with 802.11p while DYMO gives outstanding performance with 802.11.

## 1. INTRODUCTION

Mobile Ad-hoc Network (MANETs) is collection of independent mobile users taken as mobile nodes that communicate through wireless links. The creation of network protocols for these network topologies is a complex issue. Vehicular Ad-hoc Networks (VANETs) are distributed, Self-assembling communication networks that are made up of multiple autonomous moving vehicles, and peculiarized by very high node mobility. The major purpose of VANETs is providing protection and ease to the travelers.

There is no single unique protocol that is convenient for all networks impeccably. The protocols have to commensurate to network's unique characteristics, such as density, scalability and the mobility of the nodes. The routing protocols subdivided into table driven and on-demand based on the behavior of protocols. In table driven, proactive protocols are based on periodic exchange of control messages and maintaining routing tables. Each node maintains complete information about the network topology locally. However, the reactive protocol tries to discover a route only on-demand, when it is necessary. It usually takes more time to find a route compared to a proactive protocol. Our stimulation work is based upon comparison of three protocols in MANETs and VANETs named as OLSR [1], DSDV [2] and DYMO [3].

## 2. RELATED WORK AND MOTIVATION

Several papers have been published regarding the comparison of routing protocols in different simulation scenarios. The comprehensive modeling for VANETs is been done in Khabazian et.al [4-5]. This work is been improved and more generalized in section III. The comparison for AODV and DSR is carried out in realistic urban scenario with varying Node mobility and Vehicle density to observe the behavior of both protocols [6]. In this study, modified version of OLSR has been discussed. In accordance to this work we made some modifications in all other routing protocols which is shown below. The paper also shows comparison of DSR and DSDV with four different mobility models i.e., Random Waypoint, Group Mobility, Freeway and Manhattan model is presented in [7]. A few studies are carried out to evaluate the performance of different routing protocols in VANETs for some scenarios [8].

## 3. MOBILITY MODEL FOR VANETs

In this section, we will present the probability distribution function $pdf$ of the distance of a node from the strip segment. In [5], movement of each node is taken as a function of time consisting of a sequence of random intervals called mobility epochs.

The distance of a node as a function of time from the highway segment, X(t), follows a normal distribution with the $pdf$, in eq. (1).

$$b_{x(t)}(r) = \frac{1}{\sqrt{2\pi\Theta_{x(t)}}} e^{\frac{-(r-\varepsilon_{x(t)})^2}{2\Theta_{x(t)}}} \quad (1)$$

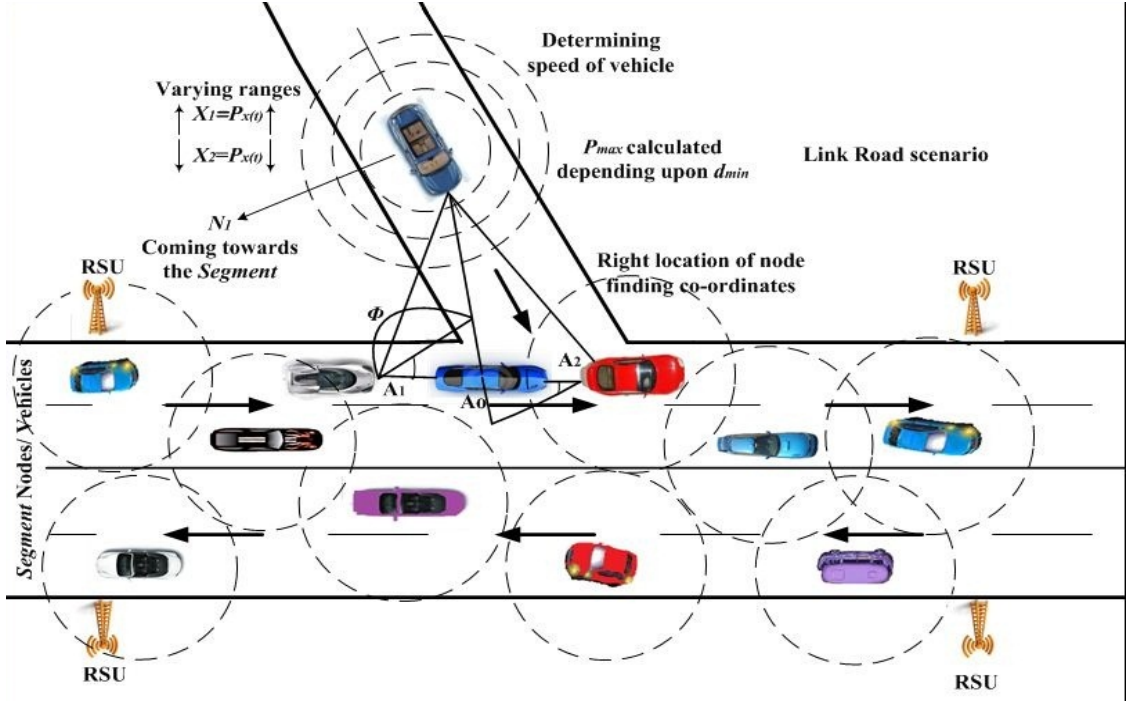

Figure 1: System Model

$$P_{x(t)}(r) = \int_0^r p_t z dz. \quad (2)$$

Taking the distance of the nodes as in [5] and simulation time written in Table[1]. We deduce that the probability $P_{xn(t)}$ of node $N1$ as shown in Fig. 1 increases as distance is greater with the strip of segments. Now calculating the efficiency of these probability, showing how much efficient is the communication between node and strip of the segment. As,

$$\eta = p_{xn(t)} * 100 \quad (3)$$

Now, $N1$ as shown in the fig.1 is coming towards the strip of the segment establishing communication. We consider $N1$ as a sender node and strip of the segment as the receiving nodes such that $N1$ sends some messages received by *Segment* nodes. Denoting the distance as $d(N1, Segment)$ between $N1$ and *Segment*.

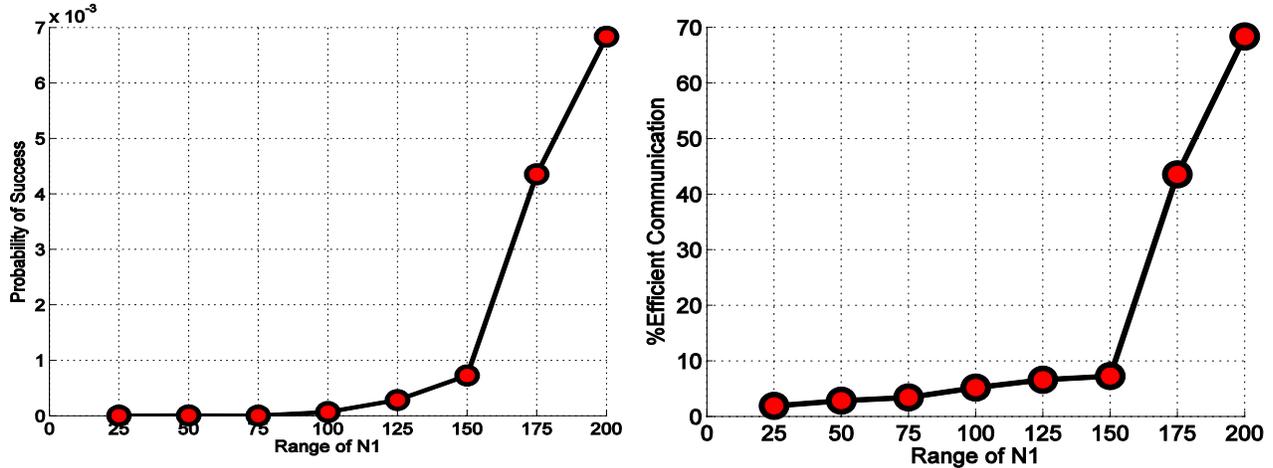

Figure 2: Probability and Efficiency of successful communication

## 4. SIMULATIONS AND DISCUSSIONS

The simulation scenario consist of Highway model involving Vehicles moving in two directions in the same way as it happens in four-lane real Highway environment. The simulations are performed with two Mac layer protocols 802.11 and 802.11p. DSDV, DYMO and OLSR are used as routing layer protocols with both Mac layer protocols. The comparison of all these protocols is done by varying the mobility and density of Vehicles. The performance metrics used are shown in Table. 1 *Throughput* is the measure of data received per unit time. Its unit is bytes per second (bps).

In Fig. 3.a maximum throughput is generated by DYMO while MOD OLSR, OLSR and DSDV. MOD DYMO shows lowest throughput and it is not performing as well as DYMO. DYMO is a reactive protocol it can only transmits or receives the routing packets when data arrives. MOD DYMO's efficiency is reduced due to the fact that number of routing packets sent per second by each node is decreased. Fig.3a having scalability scenario DYMO is on the top once again while the other protocols shows decrease throughput. MOD DYMO again shows throughput lower than that of original DYMO because DYMO is allowed to use higher bandwidth.

Nodes are mobile at a constant speed of 15*m/s* for all scalabilities scenarios. MOD OLSR and OLSR both produce good amount of throughput because of their proactive nature and ability of decision made by each node; each node decides route up to next two hops. MOD OLSR performed slightly better because it's NRL is increased by increasing the number of control packets. Increase in the number of control packets is achieved by reducing the time interval for *HELLO* and *TCmessages*. DSDV did not produce enough throughputs. It happens because the topology is varying quickly and node find a route too often to send the packets. But when we increased the time between trigger updates, periodic updates and periodic update interval these issues was resolved and more packets are sent to achieve higher throughput. In Fig. 3.b for low mobility DYMO again produce the maximum throughput while OLSR, DSDV, MOD DSDV, OLSR MOD and MOD DYMO produced in descending order. For high mobility the maximum throughput is generated by the two of proactive protocols DSDV and OLSR. MOD OLSR produces the highest amount of throughput while DSDV, OLSR and MOD DSDV also produces good amount of throughput. Due to their proactive nature and the trigger updates by DSDV and MOD DSDV and *HELLO* plus *TCmessages* allows the DSDV and OLSR to know the state of route. Due to which the packets are more likely to reach their destination.

| Parameters | Values |
|---|---|
| Simulator | NS-2(Version 2.34) |
| Channel type | Wireless |
| Radio-propagation model | Nakagami |
| Network interface type | Phy/WirelessPhy, Phy/WirelessPhyExt |
| MAC Type | Mac /802.11, Mac/802.11p |
| Interface queue Type | Queue/DropTail/PriQueue |
| Bandwidth | 2Mb |
| Packet size | 512B |
| Packet interval | 0.03s |

| Number of mobile node | 25 nodes, 50 nodes, 75 nodes,100 nodes |
|:---:|:---:|
| Speed | 2 m/s,7 m/s,15 m/s,30 m/s |
| Traffic Type | UDP, CBR |
| Simulation Time | 900 s |
| Routing Protocols | DSDV, DYMO, OLSR, MOD DSDV |

TABLE I: Simulation Parameters

In Fig. 3.c in VANETs for low scalability MOD DYMO has produced the highest throughput when used with VANETs. The DYMO depends on link layer's feedback for activation and deactivation of routes. Since 802:11*p* is better than 802:11 therefore MOD DYMO produces better throughput. MOD OLSR is slightly better than OLSR because of more proactiveness. From Fig. 3.c we can observe that after getting some improvements in MOD DSDV and in throughput of MOD OLSR. DYMO has been producing the minimum throughput because of high route timeout which means a useless route is stored for long period of time. MOD DSDV improved the performance of DSDV because of decreased robustness and proactiveness. In low mobility MOD DYMO outperforms all other protocols as observed from Fig. 3.d although DYMO is been acting decently but in the given scenario the MOD DYMO proved its worth due to its smaller route timeout interval and decreased in RREQ wait time. Due to smaller route timeout the need of finding a new route is increased which means each route is valid (possibly). Also wait time decreases means more number of RREQs that leads to increased possibility of finding a new route. DSDV is working well for low mobility because of proactive nature while MOD DSDV works reasonable but not better than DSDV. OLSR also works well but MOD OLSR is working better due to more number of *HELLO* and *TC messages*. When the case of high mobility is taken, the MOD DYMO is again performing best. While DYMO itself is 50% less efficient than MOD DYMO because of high mobility the possibility of maintaining a route for long period of time is difficult therefore DYMO is the one with fewer throughputs than MOD DYMO. OLSR and MOD OLSR both are underachievers due to very proactive nature.

*E2ED* It is the time required for a packet to reach its destination. When the scenario is for MANET and the node density is low, DSDV produces high E2ED depicted in Fig. 4.a because it is proactive protocol therefore it is not well suited for mobility. For MOD DSDV the E2ED decreases considerably. The reason is that on a preexisting path the packets are sent. Then there are only two possibilities packet is dropped or reaches its destination quickly. OLSR and MOD OLSR produced quite impressive average E2ED. It is due to the fact that it maintains a view of neighbors which stay as its neighbors for a long time or the neighbors are more stable, therefore it produces a low E2ED. DYMO has produced very impressive E2ED in the current situation. Because of the fact that it holds a route for long time and also number of control packets sent are more opposite of above conditions.

For the scenario of high scalability view from Fig. 4.a DSDV has produced a high E2ED because the routes are frequently changing and waited settling time is a problem because the routes have changed too many times. Also the less interval of trigger updates means the route can change before is reached. When the trigger interval is increased then the stable routes are found more often. OLSR gives a high E2ED because of less proactive nature while
MOD OLSR gives less delay because of more proactiveness. Overall OLSR has produced more E2ED than other routing protocols in the scalability. DYMO gives an average E2ED because of reactive nature and source routing. MOD DYMO gives low E2ED than DYMO because of simplified nature (Decreased *Route Timeout Interval* and *Rate Limit*); that is shown in fig.4.a. When the low mobility is considered the DSDV has high E2ED and the MOD DSDV is again performing well.

The Fig. 4.b reveals that for OLSR the E2ED is lower than MOD OLSR. DYMO is the best in E2ED for low mobility. For high mobility the DSDV and OLSR are not performing better than DYMO or their MOD versions. The reason for high E2ED of DSDV and OLSR is Proactive nature, Stale routes and *MPRs* loss in OLSR. DYMO is being reactive protocol have reasonable E2ED nor higher and neither low. MOD DYMO has slightly high E2ED because of the need of finding new routes more often.
For the case of low scalability Fig. 4.c shows that MOD OLSR is the best in case of E2ED while MOD DYMO and OLSR are also competing with it. OLSR, DSDV and MOD OLSR have such variation because of 802.11p. It is obvious that for high scalability the behavior of DSDV is same as in MANETs. The Fig. 4.d when simulations are performed in VANETs and variable mobility scenario then E2ED decreases by small value for OLSR and MOD OLSR but order remains the same. For DYMO, order is same only values have changed slightly. For low and high mobility behavior of all protocols is the same.

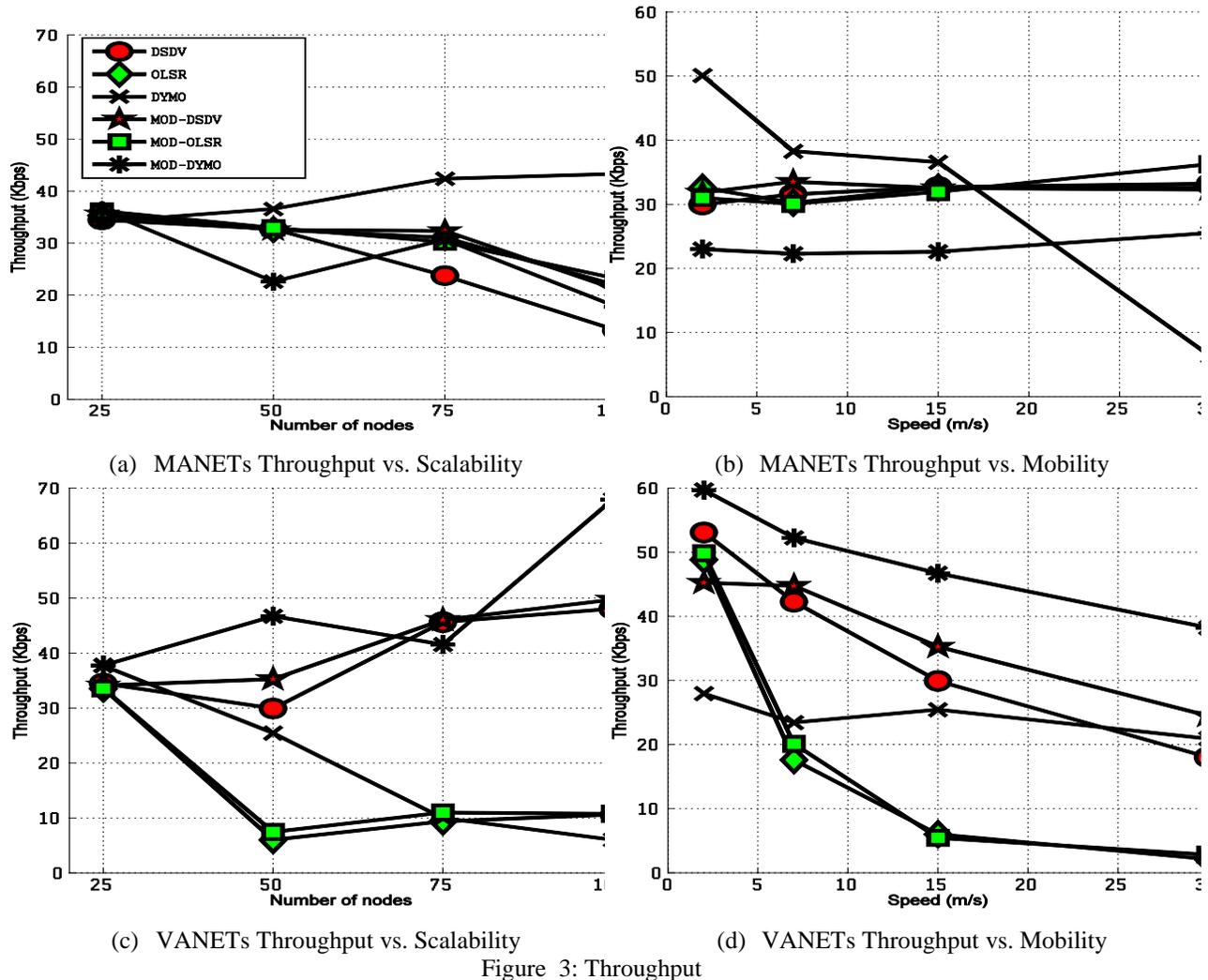

(a) MANETs Throughput vs. Scalability
(b) MANETs Throughput vs. Mobility
(c) VANETs Throughput vs. Scalability
(d) VANETs Throughput vs. Mobility

Figure 3: Throughput

*NRL* is number of control messages transmitted to receive one data packet. Fig. 5a In MANETs NRL taken in low scalability scenario. The DYMO having highest NRL while MOD DSDV and DSDV is having the lowest. When taken in higher scalability the value of DYMO relatively increases as compared to other protocols and MOD DSDV shows the same lower behavior. NRL depends on number of routing messages RERR, RREP, RREQ and *HELLOperiodicmessages*. DYMO reduces the messages so its value must be less but in comparison with the proactive protocols its value is very high. In MOD DYMO causes an increase in NRL for smaller networks. Rate limit is reduced which will eventually help in decreasing NRL of MOD DYMO as compared to DYMO but E2ED may increase. MOD DSDV NRL was expected to decrease due to the modifications that have been involved minimum time between trigger update due to the fact that less frequent topology updates.

In Fig. 5.b in MANETs, NRL is taken in mobility scenario. At low speed of nodes its value is high but at 15& 30$m=s$ its value gone decreases. MOD DYMO shows the higher value and MOD DSDV shows the very less. The overall behavior of NRL is same in both low and higher scalability. The modifications that have been made in MOD DSDV, NRL was expected to decrease due to the modifications that have been made involved minimum time between trigger update due to the fact that less frequent topology updates. The modifications have also been made in MOD DYMO also governs that for larger networks the NRL is relatively lower. Rate limit is reduced which will eventually help in decreasing NRL of MOD DYMO as compared to DYMO.

In Fig. 5.c in VANETs when NRL is taken in low scalability shows that DSDV, MOD DSVD, DYMO and MOD DYMO shows relatively lesser value as compared to OLSR and MOD OLSR. When
NRL taken in high scalability in the same scenario the MOD OLSR shows the highest value while MOD DSDV shows the very less value.

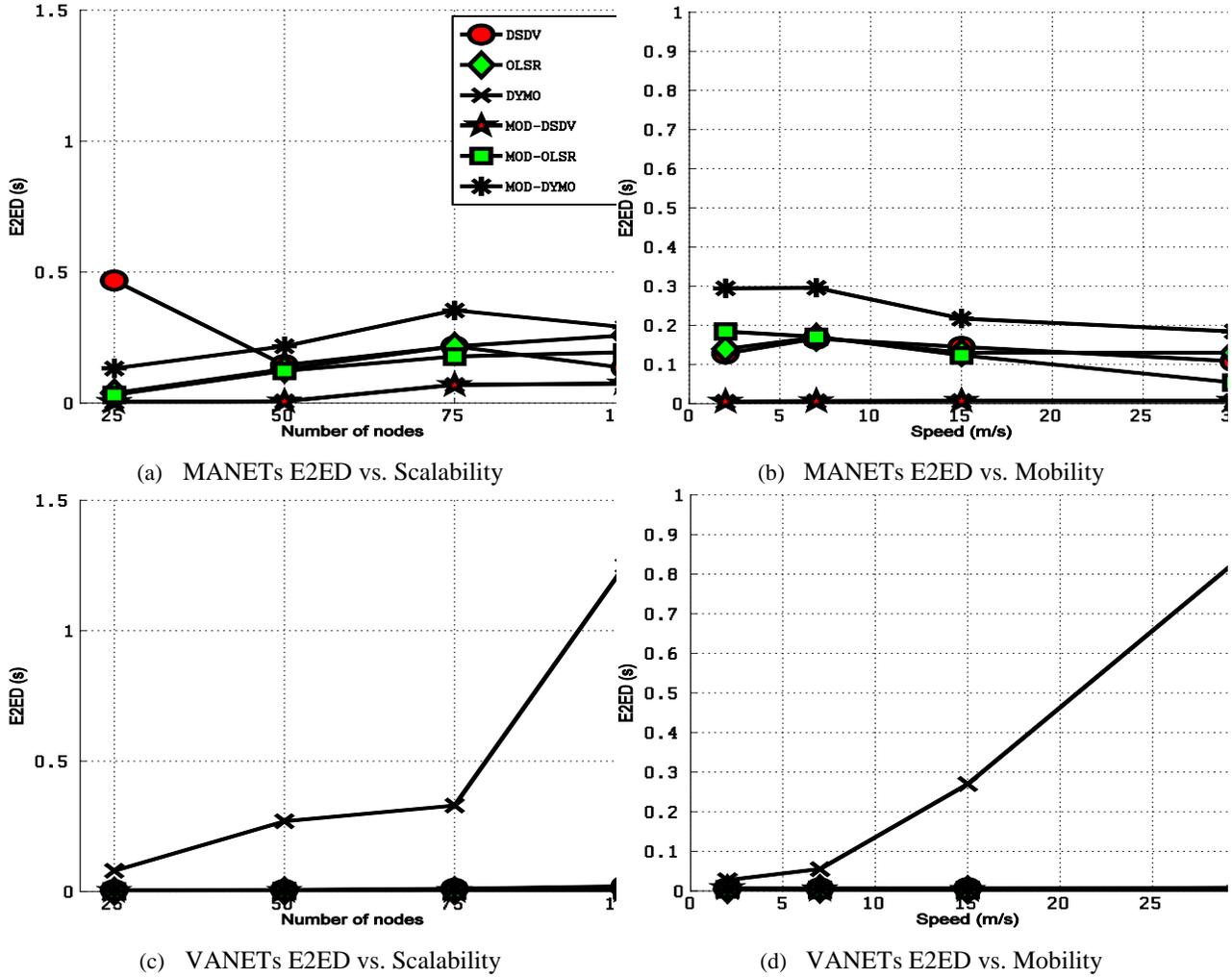

Figure 4: E2ED

From Fig. 5.c we can conclude that the normalized routing load increases to a certain extent as number of nodes increased which is expected because for more nodes more control packets are exchanged and number of MPR sets also increases. Also the nodes are moving at a constant speed of 15*m=s* which forces topological changes and also causes routing overhead. When OLSR is simulated with 802.11p. When we talk about the MOD OLSR modifications same response has been shown. HELLO messages or TC messages time interval has been decrease so that the NRL value become greater so that in less time more routing packets are flowing in the network topology.
So that it make good view of whole topology and more and more packets will reach to its destination.
In Fig. 5.d VANETs NRL is taken in mobility scenario shows that its value has no significant change with the increase in speed of nodes. DSDV shows very less value as compared to the other routing protocols in VANETs. The MOD OLSR value is very high, at low speed of nodes its value is high but at 15 and 30*m=s* its value gone decrease. As in DSDV nodes are continuously updating their routing table so with the increase in number of nodes either or mobility it's easy for their route maintenance so it shows less NRL value.

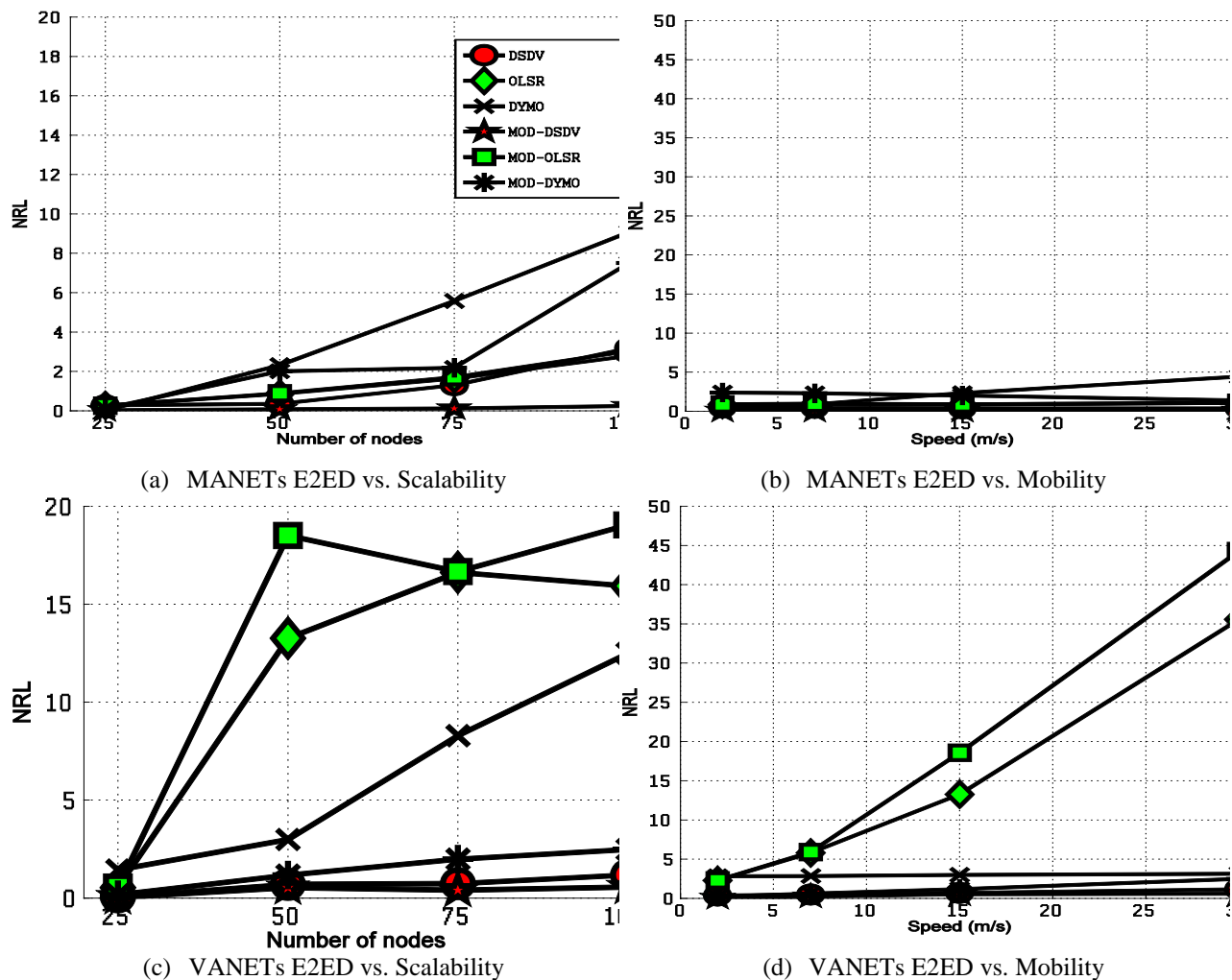

Figure 5: NRL

## 5. CONCLUSION AND FUTURE WORK

Routing protocols DSDV, DYMO and OLSR were compared for MANETs and VANETs. Our stimulation work found that overall DSDV performs fine for throughput i.e., maximum number of packets reached their destination successfully. DYMO and OLSR were underperforming and gave below throughput. DSDV and OLSR being proactive protocol stores the route as routing table entries to all destinations therefore they have the minimum E2ED while DYMO is a reactive protocol event then it worked fine when simulated with 802.11 but its performance became degraded when the Mac protocol was changed to 802.11p. When we considered the NRL, DYMO and OLSR are the ones with high NRL. Besides the evaluating the performance of DSDV, OLSR and DYMO we also made some modifications to these routing protocols and observed their performance at the end we came up with the result that with minor.

In future, we are interested to apply the same analysis on quality link metrics proposed in [9-12] and at MAC layer as [13, 14].

## REFERENCES

[1] http://hipercom.inria.fr/olsr/.
[2] Arafatur Rahman, Saiful Azad, Farhat Anwar, Aisha Hassan Abdalla "A Simulation Based Performance Analysis of Reactive Routing Protocols in Wireless Mesh Networks" International Conference on Future Networks, 2009.
[3] Charles E. Perkins and Pravin Bhagwat "Highly Dynamic Destination Sequence Distance Vector (DSDV) for Moblie Computers" 1994.

[4] Khabazian, M. and Mehmet Ali, MK "A performance modeling of vehicular ad hoc networks (VANETs)" Wireless Communications and Networking Conference, 2007. WCNC 2007.

[5] Khabazian, M. and Mehmet Ali, MK "Generalized performance modeling of vehicular Ad Hoc networks (VANETs) " Computers and Communications, 2007. ISCC 2007.

[6] Jerome Haerri, Fethi Filali, and Christian Bonnet "Performance Comparison of AODV and OLSR in VANETs Urban Environments under Realistic Mobility Patterns.

[7] Imran Khan, A "Performance Evaluation of Ad Hoc Routing Protocols For Vechicular Ad Hoc Networks" Thesis Presented to Mohammad Ali Jinnah University Fall, 2009.

[8] Moreira, W., Aguiar, E., Abelém, A., Stanton, M., "Using multiple metrics with the optimized link state routing protocol for wireless mesh networks", Simpósio Brasileiro de Redes de Computadorese Sistemas Distribuídos, Maio (2008).

[9] Usman A. etal., "An Interference and Link-Quality Aware Routing Metric for Wireless Mesh Networks,". IEEE 68th Vehicular Technology Conference, 2008.

[10] Javaid. N, Javaid. A, Khan. I. A, Djouani. K, "Performance study of ETX based wireless rout- ing metrics," 2nd IEEE International Conference on Computer, Control and Communications (IC4-2009), Karachi, Pakistan, pp.1-7, 2009.

[11] Javaid. N, Bibi, A, Djouani, K., "Interference and bandwidth adjusted ETX in wireless multi-hop networks", IEEE International Workshop on Towards Samart Communications and Network Technologies applied on Autonomous Systems (SaCoNaS2010) in conjunction with 53rd IEEE International Conference on Communications (ICC2010), Ottawa, Canada, 2010., Miami, USA, 1638-1643, 2010.

[12] Javaid. N, Ullah, M, Djouani, K., "Identifying Design Requirements for Wireless Routing Link Met- rics", 54th IEEE International Conference on Global Communications (Globecom1012), Houston, USA, pp.1-5, 2011.

[13] Dridi. K, Javaid. N, Daachi. B, Djouani. K, "IEEE 802.11 e-EDCF evaluation through MAC- layer metrics over QoS-aware mobility constraints", 7th International Conference on Advances in Mobile Computing & Multimedia (MoMM2009), Kuala Lumpur, Malaysia, 2009

[14] Dridi. K, Javaid. N, Djouani. K, Daachi. B, "Performance Study of IEEE802.11e QoS in EDCF- Contention-Based Static and Dynamic Scenarios", 16th IEEE International Conference on Electronics, Circuits, and Systems (ICECS2009), Hammamet, Tunisia, 2009.